\documentstyle[11pt,a4wide]{article}

\begingroup\makeatletter\ifx\SetFigFont\undefined
\def\x#1#2#3#4#5#6#7\relax{\def\x{#1#2#3#4#5#6}}%
\expandafter\x\fmtname xxxxxx\relax \def\y{splain}%
\ifx\x\y   % LaTeX or SliTeX?
\gdef\SetFigFont#1#2#3{%
  \ifnum #1<17\tiny\else \ifnum #1<20\small\else
  \ifnum #1<24\normalsize\else \ifnum #1<29\large\else
  \ifnum #1<34\Large\else \ifnum #1<41\LARGE\else
     \huge\fi\fi\fi\fi\fi\fi
  \csname #3\endcsname}%
\else
\gdef\SetFigFont#1#2#3{\begingroup
  \count@#1\relax \ifnum 25<\count@\count@25\fi
  \def\x{\endgroup\@setsize\SetFigFont{#2pt}}%
  \expandafter\x
    \csname \romannumeral\the\count@ pt\expandafter\endcsname
    \csname @\romannumeral\the\count@ pt\endcsname
  \csname #3\endcsname}%
\fi
\fi\endgroup
 
\makeatletter

\@addtoreset{equation}{section}
\def\section{\@startsection {section}{1}{\z@}{-3.5ex plus -1ex minus
 -.2ex}{2.3ex plus .2ex}{\large\bf\centering}}
\def\subsection{\@startsection{subsection}{2}{\z@}{-3.25ex plus%
 -1ex minus -.2ex}{1.5ex plus .2ex}{\bf}}
\def\subsubsection{\@startsection{subsubsection}{3}{\z@}{-3.25ex plus%
 -1ex minus -.2ex}{1.5ex plus .2ex}{\sl}}
\makeatother
 
\newcommand{\st}{{\rm spin}}
\newcommand{\nn}{\nonumber\\}
\newcommand{\uqgt}{\widetilde{{\cal U}_q(\hat{g})}}
\newcommand{\uqg}{{\cal U}_q({\hat g})}

\newcommand{\qda}{{\cal U}_q\left(d_4^{(3)}\right)}
\newcommand{\qgt}{{\cal U}_q(g_2)}
\newcommand{\qat}{{\cal U}_q(a_2)}
\newcommand{\qdat}{\widetilde{{\cal U}_q\left(d_4^{(3)}\right)}}
\newcommand{\qn}[1]{[ #1 ]_q}
\newcommand{\R}{{\widehat{\cal R}}(x,q)}

\newcommand{\repvect}[3]{ | {\bf #1}\ ;\ #2 ,  #3\rangle}
\newcommand{\id}{{\rm I}}
\newcommand{\ggblock}[1]{\left(#1\right)}
\begin{document}

\parindent 12pt
\parskip 9pt

{
\parskip 0pt
\newpage
\begin{titlepage}
\begin{flushright}
SISSA ref. 28/97/EP\\
Revised on 1/3/97
\end{flushright}
\vspace{2cm}
\begin{center}
{\Large{\bf The R-matrix of the $\qda$ algebra and 
$g_2^{(1)}$ affine Toda field theory
}}\\[1.5cm]
{\large
G\'abor Tak\'acs%
\footnote{On leave from Institute for Theoretical Physics, 
        E\"otv\"os University, Budapest, Hungary}%
}\\[8mm]
{\em Scuola Internazionale Superiore di Studi Avanzati (SISSA)}\\
{\em Trieste, Italy}
\\[8mm]
{27th February 1997} 
\\[8mm]
{\bf{ABSTRACT}}
\end{center}
\begin{quote}
The $R$-matrix of the $\qda$ algebra is constructed in the
$8$-dimensional 
fundamental representation. Using this result, 
an exact $S$-matrix is conjectured 
for the imaginary coupled $g_2^{(1)}$ affine Toda field theory, the 
structure of which is found to be very similar to the previously 
investigated $S$-matrix of $d_4^{(3)}$ Toda theory.  
It is shown that this $S$-matrix is consistent with the results for 
the case of real coupling using the breather-particle correspondence. 
For $q$ a root of unity it is argued that the theory can be restricted
to 
yield $\Phi(11|12)$ perturbations of $WA_2$ minimal models. 
\end{quote}
\vfill
\end{titlepage}
}

\section{Introduction}

Imaginary coupled affine Toda field theories have attracted a lot 
of interest recently (see \cite{corrigan} and references therein). 
These theories can be thought of as natural generalizations 
of sine-Gordon theory with solitonic excitations in their spectra. 
While in general these models are nonunitary as quantum field theories, 
their RSOS restrictions correspond to perturbations of W-symmetric 
rational conformal field theories (RCFTs), among them to unitary ones 
\cite{Wrestr,vaysburd}. 

In the case of theories associated to simply-laced affine Lie algebras, 
the semi-classical mass ratios are stable under quantum 
corrections \cite{hollowood}
and the $S$-matrices can be obtained using the fact 
that the theories are invariant under a quantum affine symmetry algebra 
of nonlocal charges. In the nonsimply-laced case, while the mass 
ratios are not stable under quantum corrections \cite{unstable}, 
it is again thought 
that the $S$-matrix can be obtained using the representation theory 
of the nonlocal symmetry algebra. We remark that the existing
computations 
of the mass renormalization \cite{unstable} in the nonsimply-laced case 
do not agree with each other and also that the instability of the 
classical solutions \cite{khastgir} casts a big question mark over 
the validity of the results in \cite{hollowood,unstable}. 
However, it is still plausible that the mass ratios of the
nonsimply-laced 
theories would be changed by quantum corrections similarly 
to the real coupling theories \cite{braden}, but it is unclear how to
make a 
consistent semiclassical quantisation in the imaginary coupling case. 

A nice review of the concept of applying the quantum symmetry algebra 
to construct exact $S$-matrices can be found in \cite{delius}. 
In several cases the exact $S$-matrices have been 
computed: for $a_n^{(1)}$ affine Toda theory \cite{An1smat}, for 
the $d_n^{(2)}$ \cite{Dn2smat} and the $b_n^{(1)}$ \cite{Bn1smat} case. 
Recently, an exact $S$-matrix was conjectured for the $d_4^{(3)}$ theory
as well \cite{g2}.

In this paper we will treat 
the imaginary coupled $g_2^{(1)}$ affine Toda field theory. 
This theory is known to have a $\qda$ symmetry algebra 
\cite{feldlecl,berlecl}. The fundamental excitations are solitons 
which are assigned to an $8$-dimensional irreducible representation 
of $\qda$. However, the construction of the $S$-matrix has been 
hindered by a multiplicity problem in the tensor product of 
two copies of this representation, which prevented the application 
of the usual tensor product graph (TPG) method \cite{tenspg}. In this 
paper this difficulty is circumvented by explicitely constructing the 
invariant tensors and then computing the invariant amplitudes 
using computer algebra.

The layout of the paper is the following: in Section 2 we briefly 
review the known facts about the quantum symmetry. 
The derivation of the $R$-matrix is described in Section 3. The 
S-matrix is constructed in Section 4 and the scattering amplitudes 
associated to the first breather bound state of the fundamental solitons 
are analyzed. It is shown that they can be brought into corrspondence 
with the $S$-matrix of the second particle in the real coupling theory. 
We also discuss the issues related to the gradation of the quantum
affine 
symmetry algebra and the link of the theory to 
$\Phi(11|12)$ perturbations of $WA_2$ minimal models. 
In Section 5 we draw our conclusions.

\section{The quantum affine symmetry}

Let us take an affine Lie algebra $\hat g$ and define the affine 
Toda field theory with the Lagrangian
\begin{equation}
S=\int d^2x \frac{1}{2}\partial_\mu\vec{\Phi}\partial_\mu\vec{\Phi}
+\frac{\lambda}{2\pi}\int d^2x \sum\limits_{\vec{\alpha}_j}
\exp\left(i\beta\frac{2}{(\vec{\alpha}_j,\vec{\alpha}_j)}
\vec{\alpha}_j\cdot\vec{\Phi}\right)\ ,
\label{atftlagr}\end{equation}
where the vectors $\vec{\alpha}_j\ ,\ j=0\dots r$ are the simple roots 
of $\hat g$ (meaning the simple roots of $g$ plus the extending or
affine root with label $0$). The normalization of the roots is given by
taking 
$(\vec{\alpha}_j,\vec{\alpha}_j)=2$ for the long roots.

In the usual nomenclature, (\ref{atftlagr}) is referred to as the 
${\hat g}^\vee$ affine Toda action, where ${\hat g}^\vee$ denotes 
the affine Lie algebra dual to $\hat g$, 
whose roots $\vec{\gamma}_j$ are the coroots of $\hat g$
\begin{equation}
\vec{\gamma}_j=\frac{2\vec{\alpha}_j}{(\vec{\alpha}_j,\vec{\alpha}_j)}
\ .
\end{equation} 
It is immediately apparent that any simply-laced affine Lie algebra is 
self-dual, while for nonsimply-laced ones the dual is obtained by 
reversing the arrows in the Dynkin diagram.

The theory (\ref{atftlagr}) is known to be integrable. Besides the 
infinite number of commuting charges, however, there is a nonlocal 
non-abelian symmetry algebra, which is given by the quantum symmetry 
algebra ${\cal U}_q({\hat g})$ \cite{feldlecl,berlecl}. 
The parameter $q$ is related to 
the coupling constant by 
\begin{equation}
q=\exp\left(\frac{4\pi^2 i}{\beta^2}\right)\ .
\end{equation}

To fix our conventions of the quantum affine algebra, let us 
briefly summarize the defining relations. We define first 
the algebra $\uqgt$, which is generated 
by elements $\{ h_i,\ e_i,\ f_i,\ i=0\dots r\}$, satisfying the 
following commutation relations:
\begin{eqnarray}
&&\left[h_i,h_j\right]=0,\ \left[h_i,e_j\right]=a_{ij}e_j,\ 
\left[h_i,f_j\right]=-a_{ij}f_j,\nonumber\\
&&\left[e_i,f_j\right]=\delta_{ij}\frac{q_i^{h_i}-q_i^{-h_i}}{q_i-q_i^{-1}},\ 
q_i=q^{(\alpha_i,\alpha_i)/2}\ ,
\end{eqnarray}
together with the quantum Serre relations
\begin{eqnarray}
\sum\limits_{k=0}^{1-a_{ij}}(-1)^k\left(\matrix{1-a_{ij}\cr
k}\right)_{q_i}
e_i^ke_je_i^{1-a_{ij}-k}=0\ ,\nn
\sum\limits_{k=0}^{1-a_{ij}}(-1)^k\left(\matrix{1-a_{ij}\cr
k}\right)_{q_i}
f_i^kf_jf_i^{1-a_{ij}-k}=0\ ,\nn 
i\neq j\ ,
\end{eqnarray}
where
\begin{equation}
\left(\matrix{m\cr k}\right)_{q_i}=\frac{[m]_q!}{[k]_q![m-k]_q!}\ ,\ 
[m]_q!=\prod\limits_{1\leq i\leq m}[i]_q\ ,\
[i]_q=\frac{q^i-q^{-i}}{q-q^{-1}}
\end{equation}
are the usual quantum binomial coefficients and
\begin{equation}
a_{ij}=\frac{2(\alpha_i,\alpha_j)}{(\alpha_j,\alpha_j)}
\end{equation}
is the Cartan matrix of $d_4^{(3)}$. The coproduct is given by 
\begin{eqnarray}
&&\Delta (e_i)=q_i^{-h_i/2}\otimes e_i+e_i\otimes q_i^{h_i/2}\ ,\nn
&&\Delta (f_i)=q_i^{-h_i/2}\otimes f_i+f_i\otimes q_i^{h_i/2}\ ,\nn
&&\Delta(h_i)=1 \otimes h_i + h_i \otimes 1\ .
\end{eqnarray}
The conserved charges possess a definite Lorentz spin.
Denoting the infinitesimal two-dimensional Lorentz generator by $D$ we
have
\begin{equation}
\label{derivation}
[D,e_i]= s_i e_i,~~[D,f_i]=- s_i f_i,
~~[D,H_i]=0,~~i=0,\dots,r.
\end{equation}
where $s_i$ is the Lorentz spin of $e_i$. Adjoining the operator $D$ to 
the algebra $\uqgt$ results in the full algebra $\uqg$. 

Denoting the Lorentz spin of an operator $A$ by $\st(A)$, 
$\st:~\uqg \rightarrow {\rm R}$ is a gradation of $\uqg$, which is
uniquely 
fixed by giving $s_0,\dots,s_r$. The change between the gradations can 
be performed with similarity transformations by exponentials of the 
Cartan elements $h_i$. 

For details of how the quantum affine symmetry acts on the multiparticle 
states and how this action can be used to constrain the two-particle 
S-matrix we refer the reader to \cite{delius}.

\section{The universal $R$-matrix of $\qda$ in the fundamental
representation}

\subsection{The fundamental representation of $\qda$}

The quantum symmetry of the $g_2^{(1)}$ theory is given by $\qda$. 
The Cartan matrix is
\begin{equation}
\left(\matrix{ 2 & -1 &  0 \cr 
              -3 &  2 & -1 \cr
               0 & -1 &  2 }\right)
\end{equation}
and the simple roots are given by 
\begin{eqnarray}
&&\alpha_0=(-3/\sqrt{2},-3\sqrt{3/2})\ , \nn 
&&\alpha_1=(\sqrt{2},0)\ ,\ 
\alpha_2=(-1/\sqrt{2},1/\sqrt{6})\ .
\end{eqnarray}
The length of the roots was chosen in such a way that the short roots 
have length $\sqrt{2}$, which is different from the usual normalization, 
but in this way the normalization of the coupling $\beta$ will be 
more convenient.

There are two important subalgebras of $\qda$: the generators 
$\{h_i,e_i,f_i,i=1,2\}$ form a subalgebra ${\cal A}_1$ isomorphic to 
$\qat$, while the algebra ${\cal A}_0$ generated by 
$\{h_i,e_i,f_i,i=0,1\}$ is isomorphic to ${\cal
U}_{q^3}\left(g_2\right)$.

We will assume that the fundamental solitons transform in the 
$8$-dimensional representation of the algebra, which is also the 
fundamental representation of ${\cal A}_1$. In this space, the algebra 
$\qdat$ is represented by the following matrices:
\begin{eqnarray}
&&h_0=\left(\matrix{ 
            -1 & 0 & 0 & 0 & 0 & 0 & 0 & 0 \cr
             0 &-1 & 0 & 0 & 0 & 0 & 0 & 0 \cr
             0 & 0 & 0 & 0 & 0 & 0 & 0 & 0 \cr
             0 & 0 & 0 & 0 & 0 & 0 & 0 & 0 \cr
             0 & 0 & 0 & 0 & 0 & 0 & 0 & 0 \cr
             0 & 0 & 0 & 0 & 0 & 0 & 0 & 0 \cr
             0 & 0 & 0 & 0 & 0 & 0 & 1 & 0 \cr
             0 & 0 & 0 & 0 & 0 & 0 & 0 & 1 }\right)\ ,\ 
h_1=\left(\matrix{ 
             1 & 0 & 0 & 0 & 0 & 0 & 0 & 0 \cr
             0 & 2 & 0 & 0 & 0 & 0 & 0 & 0 \cr
             0 & 0 & 1 & 0 & 0 & 0 & 0 & 0 \cr
             0 & 0 & 0 & 0 & 0 & 0 & 0 & 0 \cr
             0 & 0 & 0 & 0 & 0 & 0 & 0 & 0 \cr
             0 & 0 & 0 & 0 & 0 &-1 & 0 & 0 \cr
             0 & 0 & 0 & 0 & 0 & 0 &-2 & 0 \cr
             0 & 0 & 0 & 0 & 0 & 0 & 0 &-1 }\right)\ ,\nn 
&&h_2=\left(\matrix{ 
             1 & 0 & 0 & 0 & 0 & 0 & 0 & 0 \cr
             0 &-1 & 0 & 0 & 0 & 0 & 0 & 0 \cr
             0 & 0 &-2 & 0 & 0 & 0 & 0 & 0 \cr
             0 & 0 & 0 & 0 & 0 & 0 & 0 & 0 \cr
             0 & 0 & 0 & 0 & 0 & 0 & 0 & 0 \cr
             0 & 0 & 0 & 0 & 0 & 2 & 0 & 0 \cr
             0 & 0 & 0 & 0 & 0 & 0 & 1 & 0 \cr
             0 & 0 & 0 & 0 & 0 & 0 & 0 &-1 }\right)\ ,\ 
e_0=\left(\matrix{ 
             0 & 0 & 0 & 0 & 0 & 0 & 0 & 0 \cr
             0 & 0 & 0 & 0 & 0 & 0 & 0 & 0 \cr
             0 & 0 & 0 & 0 & 0 & 0 & 0 & 0 \cr
             0 & 0 & 0 & 0 & 0 & 0 & 0 & 0 \cr
             0 & 0 & 0 & 0 & 0 & 0 & 0 & 0 \cr
             0 & 0 & 0 & 0 & 0 & 0 & 0 & 0 \cr
             1 & 0 & 0 & 0 & 0 & 0 & 0 & 0 \cr
             0 & 1 & 0 & 0 & 0 & 0 & 0 & 0 }\right)\ ,\nn
&&e_1=\left(\matrix{ 
0 & 0 & 0 & 0 & 0 & 1 & 0 & 0 \cr
0 & 0 & 0 & \sqrt{\frac{\qn{3}\qn{1/2}}{2\qn{3/2}}} &
\sqrt{\frac{\qn{3/2}}{2\qn{1/2}}} & 0 & 0 & 0 \cr
0 & 0 & 0 & 0 & 0 & 0 & 0 & 1 \cr
0 & 0 & 0 & 0 & 0 & 0 & \sqrt{\frac{\qn{3}\qn{1/2}}{2\qn{3/2}}} & 0 \cr
0 & 0 & 0 & 0 & 0 & 0 & \sqrt{\frac{\qn{3}\qn{1/2}}{2\qn{3/2}}} & 0 \cr
0 & 0 & 0 & 0 & 0 & 0 & 0 & 0 \cr
0 & 0 & 0 & 0 & 0 & 0 & 0 & 0 \cr
0 & 0 & 0 & 0 & 0 & 0 & 0 & 0 }\right)\ ,\nn
&&e_2=\left(\matrix{ 
0 & 1 & 0 & 0 & 0 & 0 & 0 & 0 \cr 
0 & 0 & 0 & 0 & 0 & 0 & 0 & 0 \cr
0 & 0 & 0 & 0 & 0 & 0 & 0 & 0 \cr
0 & 0 & -\sqrt{\frac{\qn{3}\qn{1/2}}{2\qn{3/2}}} & 0 & 0 & 0 & 0 & 0 \cr
0 & 0 & \sqrt{\frac{\qn{3/2}}{2\qn{1/2}}} & 0 & 0 & 0 & 0 & 0 \cr
0 & 0 & 0 & -\sqrt{\frac{\qn{3}\qn{1/2}}{2\qn{3/2}}} &
\sqrt{\frac{\qn{3/2}}{2\qn{1/2}}} & 0 & 0 & 0 \cr
0 & 0 & 0 & 0 & 0 & 0 & 0 & 1 \cr
0 & 0 & 0 & 0 & 0 & 0 & 0 & 0 }\right)\ ,\nn
&&f_i=e_i^{tr}\ ,\ i=0,1,2\ ,
\end{eqnarray}
with the notation
\begin{equation}
[x]_q=\frac{q^x-q^{-x}}{q-q^{-1}}\ ,
\end{equation}
and $^{tr}$ denotes usual matrix transposition.

In the following we will use the $a_2$ homogeneous gradation, in which
case 
all the rapidity dependence is shifted to the generators with index $0$,
i.e.
\begin{equation}
\pi_{a_2}(h_i)=h_i\ ,\ 
\pi_{a_2}(e_i)=x^{\delta_{i0}}e_i\ ,\ 
\pi_{a_2}(h_i)=x^{-\delta_{i0}}f_i\ ,
\end{equation}
where $x$ is the spectral parameter (essentially the exponential of the 
rapidity; the precise correspondence will be given later). 
In this gradation, the generators 
$\{e_i,\ f_i,\ h_i,\ i=1,2\}$ are independent of the spectral parameter 
$x$ and so they still form a $\qat$ algebra.

We will solve the intertvining equation for the operator 
\begin{equation}
\R =P_{12}{\cal R}(x,q)\ ,
\end{equation}
where ${\cal R}(x,q)$ denotes the 
universal R-matrix in the tensor product of two fundamental
representations
and $x$ denotes the ratio $x_1/x_2$ of the spectral parameters in the
first and second space, respectively.

The equations for the generators $X\in {\cal A}_1$ look like
\begin{equation}
[\R,\Delta( X)]=0\ .
\end{equation} 
This means that $\R$ is a $\qat$ invariant operator in the space 
${\bf 8}\otimes {\bf 8}$. This space decomposes into irreducible 
representations under $\qgt$ in the following way
\begin{equation}
{\bf 8}\otimes {\bf 8}={\bf 1}\oplus {\bf 8} \oplus {\bf 8} 
\oplus {\bf 10} \oplus {\bf\overline{10}} \oplus {\bf 27}\ ,
\end{equation}
where we denoted irreducible representations of $\qat$ by their 
dimensions. The fact that $\bf 8$ occurs twice in the decomposition 
is the reason for which the usual tensor product graph method
\cite{tenspg} 
does not apply.

\subsection{Finding the invariants}

Since the grading is put on the long root, the $R$-matrix must 
commute with the generators of the subalgebra $\qat$ in the 
representation ${\bf 8}\otimes {\bf 8}$:

\begin{equation}
[\R , \Delta(X)]=0,\ X=h_1,h_2,e_1,e_2,f_1,f_2\ .
\label{a2inv}\end{equation}

The invariants in the tensor product space are given by the projectors 
to the irreducible components and, due to the fact that $\bf 8$ occurs 
twice, two intertwiners from one to the other and vice versa. To 
compute these invariants, one first finds the highest weight vectors 
in ${\bf 8}\otimes {\bf 8}$, i.e. the common kernel of 
$E_i=\Delta(e_i),\ i=1,2$. Let us denote the basis vectors of 
the representation $\bf 8$ (as given in the previous subsection) 
by $v_i,\ i=1...8$. Then the highest weight vectors are:
\begin{eqnarray}
&&\repvect{27}{2}{2}=v_1\otimes v_1\nn
&&\repvect{10}{3}{0}={q^{1/2}}v_1\otimes v_2-\frac{1}{q^{1/2}}v_1\otimes
v_2\nn
&&\repvect{\overline{10}}{3}{0}=-\frac{1}{q} v_1\otimes v_6+v_6\otimes
v_1\nn
&&\repvect{8_1}{1}{1}=q^{3/2}v_1\otimes v_4-\frac{1}{q^{3/2}} v_4\otimes
v_1
+\sqrt{\frac{q^2-q+1}{2q}} v_2\otimes v_6\nn
&&-\sqrt{\frac{q^2-q+1}{2q}} v_6\otimes v_2\nn
&&\repvect{8_2}{1}{1}=q^{3/2} v_1\otimes v_5+\frac{1}{q^{3/2}}
v_5\otimes v_1
-\sqrt{\frac{q^2+q+1}{2q}} v_2\otimes v_6\nn
&&-\sqrt{\frac{q^2+q+1}{2q}} v_6\otimes v_2\nn
&&\repvect{1}{0}{0}=q^2 v_1\otimes v_8+\frac{1}{q^2} v_8\otimes v_1
-q v_2\otimes v_7-\frac{1}{q} v_7\otimes v_1
-\frac{1}{q} v_3\otimes v_6+\nn
&&q v_6\otimes v_1+v_4\otimes v_4+v_5\otimes v_5\ ,
\end{eqnarray}
where $\repvect{R}{a}{b}$ denotes a vector in the representation $\bf R$ 
with Dynkin labels $a,\ b$. These vectors are orthogonal to each other 
but unnormalized. 

Using the formulae for the highest vectors, one then proceeds to build
up 
the other vectors of the representation. The structure of the 
representations is depicted in Figure 1, showing the way the different 
vectors are obtained by the action of the step operators 
$F_i=\Delta(f_i),\ i=1,2$.
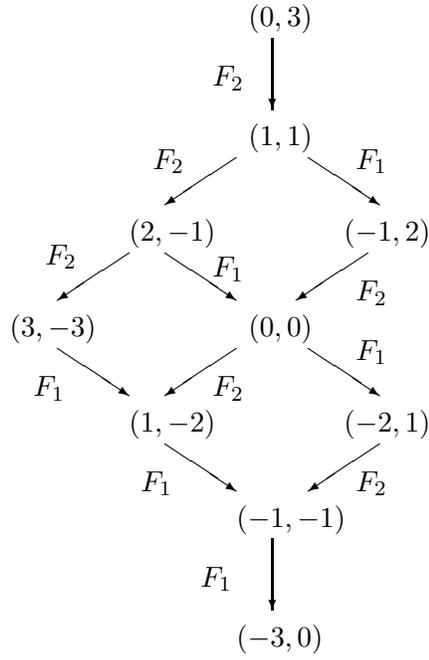
\begin{figure}
\begin{center}
\setlength{\unitlength}{0.0125in}%
\begin{picture}(110,237)(15,590)
\thinlines
\put( 95,810){\vector( 3,-2){ 30}}
\put( 30,770){\vector( 0,-1){ 30}}
\put(125,770){\vector( 0,-1){ 30}}
\put( 30,720){\vector( 0,-1){ 30}}
\put(125,720){\vector( 0,-1){ 30}}
\put(125,665){\vector(-3,-2){ 30}}
\put( 30,665){\vector( 3,-2){ 30}}
\put(110,775){\makebox(0,0)[lb]{\raisebox{0pt}[0pt][0pt]{$
(-1,2)$}}}
\put( 15,775){\makebox(0,0)[lb]{\raisebox{0pt}[0pt][0pt]{$
(2,-1)$}}}
\put( 15,725){\makebox(0,0)[lb]{\raisebox{0pt}[0pt][0pt]{$ (0,0)$}}}
\put(110,725){\makebox(0,0)[lb]{\raisebox{0pt}[0pt][0pt]{$ (0,0)$}}}
\put( 65,810){\vector(-3,-2){ 30}}
\put( 15,675){\makebox(0,0)[lb]{\raisebox{0pt}[0pt][0pt]{$
(1,-2)$}}}
\put(100,705){\makebox(0,0)[lb]{\raisebox{0pt}[0pt][0pt]{$F_1$}}}
\put(110,675){\makebox(0,0)[lb]{\raisebox{0pt}[0pt][0pt]{$
(-2,1)$}}}
\put( 60,630){\makebox(0,0)[lb]{\raisebox{0pt}[0pt][0pt]{$
(-1,-1)$}}}
\put( 70,815){\makebox(0,0)[lb]{\raisebox{0pt}[0pt][0pt]{$ (1,1)$}}}
\put(120,805){\makebox(0,0)[lb]{\raisebox{0pt}[0pt][0pt]{$F_1$}}}
\put( 25,805){\makebox(0,0)[lb]{\raisebox{0pt}[0pt][0pt]{$F_2$}}}
\put(125,645){\makebox(0,0)[lb]{\raisebox{0pt}[0pt][0pt]{$F_2$}}}
\put( 15,645){\makebox(0,0)[lb]{\raisebox{0pt}[0pt][0pt]{$F_1$}}}
\put( 80,590){\makebox(0,0)[lb]{\raisebox{0pt}[0pt][0pt]{$\bf 8$}}}
\put( 40,750){\makebox(0,0)[lb]{\raisebox{0pt}[0pt][0pt]{$F_1$}}}
\put( 40,705){\makebox(0,0)[lb]{\raisebox{0pt}[0pt][0pt]{$F_2$}}}
\put(100,750){\makebox(0,0)[lb]{\raisebox{0pt}[0pt][0pt]{$F_2$}}}
\end{picture}
\end{center}
\begin{tabular}{ccc}
\setlength{\unitlength}{0.0125in}%
\begin{picture}(150,312)(20,515)
\thinlines
\put( 80,600){\vector( 0,-1){ 30}}
\put( 95,760){\vector( 3,-2){ 30}}
\put( 35,720){\vector( 3,-2){ 30}}
\put( 95,680){\vector( 3,-2){ 30}}
\put(140,720){\vector( 3,-2){ 30}}
\put( 65,760){\vector(-3,-2){ 30}}
\put(120,720){\vector(-3,-2){ 30}}
\put( 65,680){\vector(-3,-2){ 30}}
\put(125,640){\vector(-3,-2){ 30}}
\put( 35,640){\vector( 3,-2){ 30}}
\put(170,680){\vector(-3,-2){ 30}}
\put( 70,765){\makebox(0,0)[lb]{\raisebox{0pt}[0pt][0pt]{$ (1,1)$}}}
\put( 20,725){\makebox(0,0)[lb]{\raisebox{0pt}[0pt][0pt]{$
(2,-1)$}}}
\put(110,725){\makebox(0,0)[lb]{\raisebox{0pt}[0pt][0pt]{$
(-1,2)$}}}
\put( 70,685){\makebox(0,0)[lb]{\raisebox{0pt}[0pt][0pt]{$ (0,0)$}}}
\put(155,685){\makebox(0,0)[lb]{\raisebox{0pt}[0pt][0pt]{$
(-3,3)$}}}
\put( 80,810){\vector( 0,-1){ 30}}
\put(110,645){\makebox(0,0)[lb]{\raisebox{0pt}[0pt][0pt]{$
(-2,1)$}}}
\put( 70,815){\makebox(0,0)[lb]{\raisebox{0pt}[0pt][0pt]{$ (3,0)$}}}
\put( 20,645){\makebox(0,0)[lb]{\raisebox{0pt}[0pt][0pt]{$
(1,-2)$}}}
\put( 65,605){\makebox(0,0)[lb]{\raisebox{0pt}[0pt][0pt]{$
(-1,-1)$}}}
\put( 65,555){\makebox(0,0)[lb]{\raisebox{0pt}[0pt][0pt]{$
(0,-3)$}}}
\put( 25,615){\makebox(0,0)[lb]{\raisebox{0pt}[0pt][0pt]{$F_1$}}}
\put( 25,675){\makebox(0,0)[lb]{\raisebox{0pt}[0pt][0pt]{$F_2$}}}
\put(115,615){\makebox(0,0)[lb]{\raisebox{0pt}[0pt][0pt]{$F_2$}}}
\put(115,700){\makebox(0,0)[lb]{\raisebox{0pt}[0pt][0pt]{$F_2$}}}
\put(115,675){\makebox(0,0)[lb]{\raisebox{0pt}[0pt][0pt]{$F_1$}}}
\put( 25,755){\makebox(0,0)[lb]{\raisebox{0pt}[0pt][0pt]{$F_2$}}}
\put(115,755){\makebox(0,0)[lb]{\raisebox{0pt}[0pt][0pt]{$F_1$}}}
\put(155,655){\makebox(0,0)[lb]{\raisebox{0pt}[0pt][0pt]{$F_2$}}}
\put( 55,790){\makebox(0,0)[lb]{\raisebox{0pt}[0pt][0pt]{$F_1$}}}
\put( 55,580){\makebox(0,0)[lb]{\raisebox{0pt}[0pt][0pt]{$F_2$}}}
\put(155,715){\makebox(0,0)[lb]{\raisebox{0pt}[0pt][0pt]{$F_1$}}}
\put( 25,700){\makebox(0,0)[lb]{\raisebox{0pt}[0pt][0pt]{$F_1$}}}
\put( 70,515){\makebox(0,0)[lb]{\raisebox{0pt}[0pt][0pt]{$\bf 10$}}}
\end{picture}
&\hspace{2cm}
\setlength{\unitlength}{0.0125in}%
\begin{picture}(155,322)(45,510)
\thinlines
\put(155,600){\vector( 0,-1){ 30}}
\put(170,760){\vector( 3,-2){ 30}}
\put(110,720){\vector( 3,-2){ 30}}
\put(170,680){\vector( 3,-2){ 30}}
\put(140,760){\vector(-3,-2){ 30}}
\put(195,720){\vector(-3,-2){ 30}}
\put(140,680){\vector(-3,-2){ 30}}
\put(200,640){\vector(-3,-2){ 30}}
\put(110,640){\vector( 3,-2){ 30}}
\put( 95,720){\vector(-3,-2){ 30}}
\put( 65,680){\vector( 3,-2){ 30}}
\put(145,765){\makebox(0,0)[lb]{\raisebox{0pt}[0pt][0pt]{$ (1,1)$}}}
\put( 95,725){\makebox(0,0)[lb]{\raisebox{0pt}[0pt][0pt]{$
(2,-1)$}}}
\put(185,725){\makebox(0,0)[lb]{\raisebox{0pt}[0pt][0pt]{$
(-1,2)$}}}
\put(145,685){\makebox(0,0)[lb]{\raisebox{0pt}[0pt][0pt]{$ (0,0)$}}}
\put(185,645){\makebox(0,0)[lb]{\raisebox{0pt}[0pt][0pt]{$
(-2,1)$}}}
\put(155,810){\vector( 0,-1){ 30}}
\put( 95,645){\makebox(0,0)[lb]{\raisebox{0pt}[0pt][0pt]{$
(1,-2)$}}}
\put(145,510){\makebox(0,0)[lb]{\raisebox{0pt}[0pt][0pt]{$\bf\overline{10}$}}}
\put(140,605){\makebox(0,0)[lb]{\raisebox{0pt}[0pt][0pt]{$
(-1,-1)$}}}
\put(190,700){\makebox(0,0)[lb]{\raisebox{0pt}[0pt][0pt]{$F_2$}}}
\put(190,675){\makebox(0,0)[lb]{\raisebox{0pt}[0pt][0pt]{$F_1$}}}
\put(190,755){\makebox(0,0)[lb]{\raisebox{0pt}[0pt][0pt]{$F_1$}}}
\put( 45,685){\makebox(0,0)[lb]{\raisebox{0pt}[0pt][0pt]{$
(3,-3)$}}}
\put(130,660){\makebox(0,0)[lb]{\raisebox{0pt}[0pt][0pt]{$F_2$}}}
\put(130,710){\makebox(0,0)[lb]{\raisebox{0pt}[0pt][0pt]{$F_1$}}}
\put(145,815){\makebox(0,0)[lb]{\raisebox{0pt}[0pt][0pt]{$ (0,3)$}}}
\put(140,555){\makebox(0,0)[lb]{\raisebox{0pt}[0pt][0pt]{$
(-3,0)$}}}
\put(130,790){\makebox(0,0)[lb]{\raisebox{0pt}[0pt][0pt]{$F_2$}}}
\put(125,580){\makebox(0,0)[lb]{\raisebox{0pt}[0pt][0pt]{$F_1$}}}
\put(105,755){\makebox(0,0)[lb]{\raisebox{0pt}[0pt][0pt]{$F_2$}}}
\put( 60,715){\makebox(0,0)[lb]{\raisebox{0pt}[0pt][0pt]{$F_2$}}}
\put( 55,660){\makebox(0,0)[lb]{\raisebox{0pt}[0pt][0pt]{$F_1$}}}
\put(100,620){\makebox(0,0)[lb]{\raisebox{0pt}[0pt][0pt]{$F_1$}}}
\put(190,620){\makebox(0,0)[lb]{\raisebox{0pt}[0pt][0pt]{$F_2$}}}
\end{picture}
\end{tabular}
\caption{The structure of the representations $\bf 8$, $\bf 10$ and
$\bf\overline{10}$.}
\end{figure}
The only point where one has to be careful that the two $(0,0)$ vectors
in 
$\bf 8$ are not orthogonal, so an orthogonalization must be performed in 
this subspace. Using this information, it is possible to construct the
basis 
of all irreducible components except $\bf 27$ and to compute the
invariant 
projectors ${\cal P}_{\bf R}$ on the irreducible subspaces
${\bf R}={\bf 1},\ {\bf 8_1},\ {\bf 8_2},\ {\bf 10},\ {\bf
\overline{10}}$. 
The projector ${\cal P}_{\bf 27}$ on $\bf 27$ was computed by
subtracting 
the sum of the other projectors from identity.

The two intertwiners ${\cal I}_{12}$ and ${\cal I}_{21}$ between 
$\bf 8_1$ and $\bf 8_2$ can be computed in the following way:
\begin{equation}
{\cal I}_{ij}=\sum\limits_{(n_1,n_2)} 
|{\bf 8_i}\ ;\ n_1,n_2\rangle\otimes\langle{\bf 8_j}\ ;\ n_1,n_2|\ ,
\end{equation}
where it is understood that the sum runs over an orthonormalized basis 
composed of basis vectors with Dynkin labels $n_1,\ n_2$ (the degeneracy 
indices in the $(0,0)$ subspace are not shown explicitely, but should 
be understood). It is also necessary to 
note that in all scalar products the variable $q$ must be treated as a 
formal one and should not be conjugated when computing the conjugate
vector. 
This is equivalent to the requirement that the generators of the quantum 
affine algebra obey the hermiticity condition
\begin{equation}
h_i^\dagger=h_i,\ e_i^\dagger=f_i\ .
\end{equation} 
\subsection{The fundamental $R$-matrix}

Now let us introduce the spectral parameter. The universal 
$R$-matrix in the $a_2$ homogeneous gradation satisfies the following 
equations in addition to (\ref{a2inv}):
\begin{eqnarray}
&&[\R,\Delta(h_0)]=0,\nn
&&\R(q^{-3h_0/2}\otimes e_0+xe_0\otimes q^{3h_0/2})=
(xq^{-3h_0/2}\otimes e_0+e_0\otimes q^{3h_0/2})\R\ ,\nn 
&&\R(q^{-3h_0/2}\otimes f_0+\frac{1}{x}f_0\otimes q^{3h_0/2})=
(\frac{1}{x}q^{-3h_0/2}\otimes f_0+f_0\otimes q^{3h_0/2})\R\ .\nn
\label{jimbo}\end{eqnarray}
We write the solution in the following form
\begin{eqnarray}
\R=&&\sum\limits_{{\bf R}={\bf 1},{\bf 10},{\bf\overline{10}},\bf{27}}
A^{\bf R}(x,q){\cal P}_{\bf R}+A^{\bf 8}_{11}(x,q){\cal P}_{\bf 8_1}+
A^{\bf 8}_{22}(x,q){\cal P}_{\bf 8_2}+\nn
&&A^{\bf 8}_{12}(x,q){\cal I}_{12}+A^{\bf 8}_{21}(x,q){\cal I}_{21}\ .
\end{eqnarray}
Due to the relation $h_0=-(2h_1+h_2)/3$, the first of the equations 
(\ref{jimbo}) 
is satisfied automatically. The third one follows from the second one 
if $\R$ is a symmetric matrix (which turns out to be the case). 
The remaining equation can be solved using the computer algebra program 
MAPLE. The result is
\begin{eqnarray}
&&A^{\bf 27}(x,q)=1\ ,\nn
&&A^{\bf 10}(x,q)=A^{\bf\overline{10}}(x,q)=\frac{1-xq^2}{x-q^2}\ ,\nn
&&A^{\bf 1}(x,q)=\frac{1-xq^2}{x-q^2}\frac{1-xq^6}{x-q^6}\ ,\nn
&&A^{\bf 8}_{11}(x,q)=
\frac{2q^5(1-x^3)-(q^6-1)((q^4+q^3-q^2-q+1)x^2+(q^4-q^3-q^2+q+1)x)}
{2(x-q^2)(x^2+xq^4+q^8)}\ ,\nn
&&A^{\bf 8}_{22}(x,q)=
\frac{2q^5(x^3-1)-(q^6-1)((q^4-q^3-q^2+q+1)x^2+(q^4+q^3-q^2-q+1)x)}
{2(x-q^2)(x^2+xq^4+q^8)}\ ,\nn
&&A^{\bf 8}_{12}(x,q)=A^{\bf 1}_{21}(x,q)=
\frac{x(x-1)(q^6-1)\sqrt{q^8+q^6+q^4+q^2+1}}
{2(x-q^2)(x^2+xq^4+q^8)}\ .
\label{ampl}\end{eqnarray}
The coefficient of ${\cal P}_{\bf 27}$ was normalized to one, using the 
fact that the $R$-matrix is only determined up to an arbitrary scalar 
factor. The above solution for $\R$ satisfies
\begin{eqnarray}
&&\R{\widehat{\cal R}}(1/x,q)=\id\ ,\label{unitarity}\\
&&{\widehat {\cal R}}(q^6/x,q) \frac{(x-1)(x-q^4)(x^2+xq^2+q^4)q^4}
{(x-q^6)(x-q^2)(x^2+xq^4+q^8)} \nn
&& =(C\otimes\id)(P_{12}\R)^{t_1}(C\otimes\id)P_{12}\
,\label{crossing}\\
&&C=\left(\matrix{
0 & 0 & 0 & 0 & 0 & 0 & 0 & q^2 \cr
0 & 0 & 0 & 0 & 0 & 0 & -1/q & 0 \cr
0 & 0 & 0 & 0 & 0 & -q  & 0 & 0 \cr
0 & 0 & 0 & 1 & 0 & 0 & 0 & 0 \cr
0 & 0 & 0 & 0 & 1 & 0 & 0 & 0 \cr
0 & 0 & -1/q & 0 & 0 & 0 & 0 & 0\cr
0 & -q  & 0 & 0 & 0 & 0 & 0 & 0\cr
1/q^2 & 0 & 0 & 0 & 0 & 0 & 0 & 0\cr
}\right)\ .\label{chconj}
\end{eqnarray}
(\ref{unitarity}) is related to the unitarity of the scattering
amplitudes, 
while (\ref{crossing}) describes how the $R$-matrix transforms under 
crossing symmetry, with (\ref{chconj}) as the charge conjugation matrix.

The $R$-matrix has the following pole structure:
\begin{itemize}
\item{} $x=q^6$: the $R$-matrix degenerates to a one-dimensional 
projector in the $\bf 1$ channel. This should correspond to breathers
occuring 
as bound states of fundamental solitons.
\item{} $x=\omega q^4$ and $x=\omega^{-1} q^4$: the $R$-matrix
degenerates to 
an eight-dimensional projector in a combination of the $\bf 8_1$ and 
$\bf 8_2$ channel. This corresponds to higher solitons in the $\bf 8$ 
representation and also includes the fundamental soliton occurring as a 
bound state of itself.
\item{} $x=q^2$: the $R$-matrix degenrates to a projector onto the 
reducible ${\bf 1}+{\bf 8}+{\bf 10}+{\bf\overline{10}}$ representation, 
where the $\bf 8$ is another combination of $\bf 8_1$ and $\bf 8_2$.
This corresponds to a new solitonic multiplet. 
\end{itemize}
The above result can be understood in the following way. The bootstrap 
structure corresponds to the $\qat$ subalgebra embedded in $\qda$. 
But the algebra $d_4^{(3)}$ is obtained by twisting $d_4^{(1)}$ 
using its $Z_3$ symmetry. The corresponding embedding of $a_2$ 
in $d_4^{(1)}$ is given by the nodes $4$ and $0$ in 
the Dynkin diagram. This embedding can be identified with a 
singular embedding of $a_2$ into $d_4$ \cite{slansky}.

Under this embedding, the branching rules are the following: 
the three $8$-dimensional fundamental representations of $d_4$ 
(corresponding to the nodes $1$, $2$ and $3$)
become the $8$-dimensional adjoint representation of $a_2$. 
The fourth fundamental representation of $d_4$ (node $4$), which is 
just the $28$-dimensional adjoint representation of $so(8)$, 
decomposes as ${\bf 8}+{\bf 10}+{\bf\overline{10}}$. 
\begin{figure}[h]
\begin{center}
\setlength{\unitlength}{0.0125in}%
\begin{picture}(380,110)(5,715)
\thinlines
\put( 60,780){\circle{10}}
\put(100,750){\circle{10}}
\put( 20,750){\circle{10}}
\put(140,820){\circle{10}}
\put(220,820){\circle{10}}
\put(280,790){\circle{10}}
\put(380,790){\circle{10}}
\put(330,790){\circle{10}}
\put(140,740){\circle{10}}
\put(220,740){\circle{10}}
\put(179,781){\circle{10}}
\put( 60,785){\line( 0, 1){ 30}}
\put( 65,780){\line( 6,-5){ 30}}
\put( 55,780){\line(-6,-5){ 30}}
\put(285,790){\line( 1, 0){ 40}}
\put(330,795){\line( 1, 0){ 50}}
\put(335,790){\line( 1, 0){ 40}}
\put(330,785){\line( 1, 0){ 50}}
\put( 60,820){\circle{10}}
\multiput(355,790)(0.40000,0.40000){26}{\makebox(0.4444,0.6667){\rm
.}}
\put(125,735){\makebox(0,0)[lb]{\raisebox{0pt}[0pt][0pt]{$ 1$}}}
\multiput(355,790)(0.40000,-0.40000){26}{\makebox(0.4444,0.6667){\rm
.}}
\put(185,785){\line( 1, 1){ 30}}
\put(185,775){\line( 1,-1){ 30}}
\put(145,745){\line( 1, 1){ 30}}
\put(145,815){\line( 1,-1){ 30}}
\put( 45,815){\makebox(0,0)[lb]{\raisebox{0pt}[0pt][0pt]{$ 3$}}}
\put( 45,785){\makebox(0,0)[lb]{\raisebox{0pt}[0pt][0pt]{$ 4$}}}
\put(  5,745){\makebox(0,0)[lb]{\raisebox{0pt}[0pt][0pt]{$ 1$}}}
\put( 80,745){\makebox(0,0)[lb]{\raisebox{0pt}[0pt][0pt]{$ 2$}}}
\put(125,815){\makebox(0,0)[lb]{\raisebox{0pt}[0pt][0pt]{$ 3$}}}
\put(235,815){\makebox(0,0)[lb]{\raisebox{0pt}[0pt][0pt]{$ 0$}}}
\put(275,765){\makebox(0,0)[lb]{\raisebox{0pt}[0pt][0pt]{$ 0$}}}
\put(330,765){\makebox(0,0)[lb]{\raisebox{0pt}[0pt][0pt]{$ 4$}}}
\put( 50,715){\makebox(0,0)[lb]{\raisebox{0pt}[0pt][0pt]{$d_4$}}}
\put(175,715){\makebox(0,0)[lb]{\raisebox{0pt}[0pt][0pt]{$d_4^{(1)}$}}}
\put(325,715){\makebox(0,0)[lb]{\raisebox{0pt}[0pt][0pt]{$d_4^{(3)}$}}}
\put(160,775){\makebox(0,0)[lb]{\raisebox{0pt}[0pt][0pt]{$ 4$}}}
\put(235,735){\makebox(0,0)[lb]{\raisebox{0pt}[0pt][0pt]{$ 2$}}}
\end{picture}
\end{center}
\caption{Dynkin diagrams of the algebras $d_4$, $d_4^{(1)}$ and
$d_4^{(3)}$}
\end{figure}
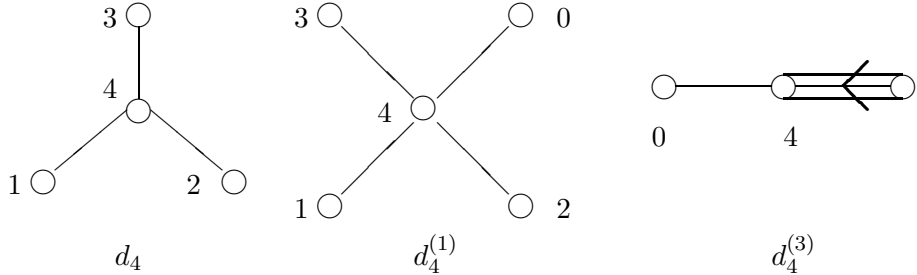
After the $Z_3$ twisting, the three $8$-dimensional representations
collapse to the $8$-dimensional 
fundamental representation of $d_4^{(3)}$ corresponding 
to the fundamental soliton. The coefficient of the pole at $x=q^2$ just 
projects onto the adjoint+singlet representation, where by adjoint we
mean 
the reducible representation  ${\bf 8}+{\bf 10}+{\bf\overline{10}}$, as 
justified by the above discussion. This is strikingly similar to the 
pattern which emerges in the case of the $R$-matrix of the 
${\cal U}_q\left(g_2^{(1)}\right)$ algebra \cite{g2}. 
Note that the adjoint representation at the quantum level is 
extended to include a singlet particle, which is again similar to 
the case of ${\cal U}_q\left(g_2^{(1)}\right)$.

The result (\ref{ampl}) shows that apart from the subspace containing 
$\bf 8_1$ and $\bf 8_2$ 
the $R$-matrix can be obtained by using the tensor product graph (TPG)
method 
\cite{tenspg}, which works well in the case when there are no
multiplicities 
in the tensor product of the representations. This is not the case here, 
but in the ${\bf 1}+{\bf 10}+{\bf\overline{10}}+{\bf 27}$ subspace the
TPG 
works as it indeed should.

One can also compute the limit of the universal $R$-matrix as 
$x\rightarrow 0$ or $\infty$:

\begin{equation}
{\cal R}(0,q)={\cal R}(\infty ,q)^{-1}=
{\cal P}_{\bf 27}-q^2{\cal P}_{\bf 10}-q^2{\cal P}_{\bf\overline{10}}-
q^5{\cal P}_{\bf 8_1}+q^5{\cal P}_{\bf 8_2}+q^8{\cal P}_{\bf 1}\ ,
\end{equation}
which coincides with the result in \cite{ma}, where a solution to 
the Yang-Baxter equation was computed in the ${\bf 8}\otimes{\bf 8}$ 
representation of $\qat$. Note that in this limit the mixing terms 
in the ${\bf 8_1}+{\bf 8_2}$ subspace disappear. 

\section{The $\qda$ invariant S-matrix}

\subsection{The scalar factor and the pole structure}

The $S$-matrix in the $a_2$ homogenous gradation takes the form
\begin{equation}
S(\theta )=\R S_0(\theta ).
\end{equation}
$S_0(\theta )$ is a scalar function of the rapidity $\theta$ to be 
determined and we have the following relation between 
$x,\ q$ and the physical parameters $\theta,\ \beta$:
\begin{equation}
q=\exp\left(\frac{4\pi^2 i}{\beta^2}\right)\ ,\ 
x=\exp\left(\left(\frac{4\pi^2}{\beta^2}h-h^\vee\right)\theta\right)\ ,
\label{xtheta}\end{equation}
where $h=6$ and $h^\vee=4$ denote the Coxeter and dual Coxeter number 
of $d_4^{(3)}$, respectively. In principle, the relation for $x$ may 
contain a phase factor \cite{Bn1smat}, but the above assignment turns
out 
to be consistent with the particle-breather correspondence using the 
bootstrap.  We introduce a new parametrization of the coupling constant
with 
\begin{equation}
\xi =\frac{\pi\beta^2}{12\pi-2\beta^2}\ ,
\end{equation}
so that $x$ and $q$ take the form
\begin{equation}
x=\exp\left(\frac{2\pi}{\xi}\theta\right)\ ,\ 
q=\exp\left(\frac{i\pi}{3}\left(\frac{\pi}{\xi}+2\right)\right)\ .
\end{equation}
In order to make the $S$-matrix crossing symmetric and unitary, 
the function $S_0(\theta )$ must satisfy
\begin{eqnarray}
S_0(i\pi-\theta)=
&&\frac{\sinh\frac{\pi}{\xi}\theta 
\sinh\frac{\pi}{\xi}\left(\theta-i\frac{\pi}{3}\right)
\sinh\frac{\pi}{\xi}\left(\theta-i\frac{\pi}{3}-i\frac{\xi}{3}\right)
}
{\sinh\frac{\pi}{\xi}\left(\theta-i\pi\right) 
\sinh\frac{\pi}{\xi}\left(\theta-i\frac{2\pi}{3}\right)
\sinh\frac{\pi}{\xi}\left(\theta-i\frac{2\pi}{3}+i\frac{\xi}{3}\right)
}\times\nn
&&\frac{\sinh\frac{\pi}{\xi}\left(\theta-i\frac{2\pi}{3}-i\frac{\xi}{3}\right)}
{\sinh\frac{\pi}{\xi}\left(\theta-i\frac{\pi}{3}+i\frac{\xi}{3}\right)}
S_0(\theta )\ .
\end{eqnarray}
This equation has many solutions, among which we choose the so-called
minimal 
one with the minimal number of poles in the physical strip 
$0\leq\theta<i\pi$. This solution is characterized by the property 
that it only has poles in 
locations where the $S$-matrix degenerates to some projector onto a
proper 
subspace of the two-particle space and is given by the following
formula:
\begin{eqnarray}
S_0(\theta )=
\prod _{k=0}^{\infty }
&&\frac
{\ggblock{1} \ggblock{\frac{2\pi}{\xi}} 
\ggblock{\frac{\pi}{3\xi}}\ggblock{1+\frac{5\pi}{3\xi}}
\ggblock{\frac{1}{3}+\frac{2\pi}{3\xi}}\ggblock{\frac{2}{3}+\frac{4\pi}{3\xi}}
}
{\ggblock{1+\frac{\pi}{\xi}} \ggblock{\frac{\pi}{\xi}} 
\ggblock{\frac{4\pi}{3\xi}} \ggblock{1+\frac{2\pi}{3\xi}}
\ggblock{\frac{2}{3}+\frac{\pi}{3\xi}}\ggblock{\frac{1}{3}+\frac{5\pi}{3\xi}}
}\times\nn
&&\frac
{\ggblock{\frac{1}{3}+\frac{\pi}{3\xi}}\ggblock{\frac{2}{3}+\frac{5\pi}{3\xi}}}
{\ggblock{\frac{2}{3}+\frac{2\pi}{3\xi}}\ggblock{\frac{1}{3}+\frac{4\pi}{3\xi}}}\
,
\label{S0prod}\end{eqnarray} 
where
\begin{equation}
\left(x\right)=
\frac{\Gamma\left(x+\frac{2k\pi}{\xi}+\frac{i\theta}{\xi}\right)}
{\Gamma\left(x+\frac{2k\pi}{\xi}-\frac{i\theta}{\xi}\right)}\ .
\end{equation}
The pole structure of the $S$-matrix can be computed using the
singularity 
structure of the $R$-matrix and the formula for $S_0(\theta )$. It can 
be seen that the poles come in crossing symmetric pairs and that the 
$S$-matrix degenerates to a projector at the direct channel pole and 
to the complementary projector at the crossed pole. Here we list 
the direct channel poles only:
\begin{itemize}
\item{} $\theta=i\pi-m\xi\ ,\ m\geq 1$: singlet bound states of the 
fundamental soliton, so-called breathers. 
We denote the $m=1$ case by $B_1$ and the 
rest for $m>1$ with $B_1^{(m-1)}$, considered to be excited states of 
$B_1$, which is the ground state of the bound system of two solitons.
\item{} $\theta=\frac{2i\pi}{3}-m\xi\ ,\ m\geq 0$: for $m=0$ this is 
the pole for the fundamental soliton $K_1$ to occur as a bound state of
two 
$K_1$ particles. The higher cases are denoted by $K_m$ and can be
thought 
of as excited states of the fundamental soliton.
\item{} $\theta=\frac{2i\pi}{3}-\left(m-\frac{2}{3}\right)\xi\ ,\ m\geq
1$: 
another series of solitons in the fundamental representation, denoted by
$L_m$.
\item{} $\theta=\frac{i\pi}{3}-\left(m-\frac{2}{3}\right)\xi\ ,\ m\geq
1$: 
a series of particles of the adjoint+singlet type (see discussion 
in subsection 3.3), which we denote by $A_m$.
\end{itemize}
Of course, the bootstrap structure turns out to be much more complicated 
since new poles can arise from amplitudes involving higher particles. We 
do not enter the question of closing the bootstrap here. The number of 
higher states strongly depends on the value of $\xi$, since only the 
poles falling into the physical strip can be candidates for new 
particles. In what follows we give an analysis of the 
amplitudes involving $K_1$ and $B_1$, which is useful for a comparison 
to the real coupling case via the breather-particle correspondence.

\subsection{Breather-soliton and breather-breather scattering
amplitudes}

To get the $S$-matrix corresponding to the breather $B_1$, first 
one needs an equation for the bootstrap of the tensor part of the 
fundamental soliton $S$-matrix. This formula is the following:
\begin{eqnarray}
&&\left({\cal P}_{\bf 1}\right)_{12}
{\cal R}_{13}(xq^2,q){\cal R}_{23}(x/q^2,q)
\left({\cal P}_{\bf 1}\right)_{12}=\nn
&&\frac{(xq^{3}-1)(x-q)(x^2q^2+xq+1)}{(x-q^3)(xq-1)(x^2+qx+q^2)}
\left({\cal P}_{\bf 1}\right)_{12}\otimes \id
\ ,\label{brfuse}
\end{eqnarray}
with ${\cal P}_{\bf 1}$ denoting the projector onto the singlet in the
tensor 
product ${\bf 8}\otimes{\bf 8}$ as before and the indices $1,2,3$
labelling 
the three one-particle spaces. Using formula (\ref{S0prod}) for the 
scalar factor $S_0(\theta )$, one finds the following results:
\begin{eqnarray}
&&S_{K_1B_1}=\left\langle \frac{\pi}{2}+\frac{\xi}{2}\right\rangle_{K_1}
\left\langle \frac{5\pi}{6}-\frac{\xi}{2}\right\rangle_{K_2}
\left\langle \frac{\pi}{6}+\frac{\xi}{6}\right\rangle
\left\langle -\frac{\pi}{6}+\frac{\xi}{6}\right\rangle_{CDD_1}\nn
&&S_{B_1B_1}=\left\langle \xi\right\rangle_{B_1^{(1)}}
\left\langle \frac{2\pi}{3}\right\rangle_{B_1}
\left\langle -\frac{\pi}{3}+\xi \right\rangle \times\nn
&&\left\langle \frac{\pi}{3}-\frac{\xi}{3}\right\rangle
\left\langle \frac{2\pi}{3}-\frac{\xi}{3}\right\rangle
\left\langle -\frac{\pi}{3}-\frac{2\xi}{3}\right\rangle_{CDD2}
\left\langle -\frac{\pi}{3}+\frac{2\xi}{3}\right\rangle\ .
\label{brS}\end{eqnarray}
where we used the notation
\begin{equation}
\langle x \rangle =\frac{\sinh\theta+i\sin x}{\sinh\theta-i\sin x}\ .
\end{equation}
The labels at the bottom of the blocks denote the bound states to which 
the corresponding poles belong. Blocks with no labels mean that there
should 
be either a new particle associated to the pole, or there is some kind
of 
multiparticle scattering process responsible for the singularity via a 
Coleman-Thun mechanism \cite{CT}. The poles labelled with $CDD_1$ and
$CDD_2$ 
are always outside the physical strip whenever $B_1$ exists (i.e. 
for $\xi<\pi$).

The $S$-matrix of the real coupled $g_2^{(1)}$ Toda theory was computed 
using the hypothesis of `floating masses' and strong--weak
coupling duality \cite{floating,duality1,duality2}.
Using the floating Coxeter number $H$, the S-matrix of the second
particle 
reads \cite{corrigan}
\begin{equation}
S_{22}=\left\langle \frac{2\pi}{H}\right\rangle
\left\langle \frac{2\pi}{3}\right\rangle
\left\langle \frac{2\pi}{H}+\frac{\pi}{3} \right\rangle
\left\langle \frac{\pi}{3}-\frac{4\pi}{H}\right\rangle
\left\langle \frac{\pi}{3}-\frac{6\pi}{H}\right\rangle
\left\langle \frac{4\pi}{H}-\pi\right\rangle
\left\langle \frac{6\pi}{H}-\pi\right\rangle \ ,
\label{realS}
\end{equation}
where
\begin{equation}
H=12\frac{6\pi+\beta'^2}{12\pi+\beta'^2}\ ,
\end{equation}
and $\beta'$ is the (real) coupling constant. 
Substituting $\beta=i\beta'$ we obtain the relation
\begin{equation}
\xi=\pi\left(\frac{6}{H}-1\right)\ ,
\end{equation}
and one can easily check that $S_{B_1B_1}$ and $S_{22}$ become
identical.

Now we face a puzzle: where is the first particle of the real coupling 
theory? To understand this let us recall the case of the $d_4^{(3)}$ 
theory with symmetry algebra ${\cal U}_q\left(g_2^{(1)}\right)$
\cite{g2}. 
There we have found two types of solitons: one in the $7$-dimensional 
fundamental representation of $g_2$ and another one in the adjoint 
representation (extended by a singlet). The first particle of the 
real coupling theory was shown to correspond to the breather $B_1$
originating 
from the fundamental soliton, while the second one corresponds to a
particle 
$AB_1$ which was conjectured to be a singlet bound state (i.e. a
breather) of 
the soliton in the adjoint representation. 

Using this as an analogy, we expect that in the case of $g_2^{(1)}$ Toda 
field theory the first particle has to come from the other type of
solitons, 
the one in the ${\bf 1}+{\bf 8}+{\bf 10}+{\bf\overline{10}}$ 
representation. Unfortunately, due to the reducibility of this 
representation the tensor product graph method is not applicable 
and the large number of dimensions prevents us from using a brute 
force approach with computer algebra. Therefore we can regard the 
statement on the origin of the `missing' particle as a conjecture that 
can possibly be verified in the future with some other approach 
which makes the computation feasible.

\subsection{Remarks on $\Phi(11|12)$ perturbations of $WA_2$ minimal 
models}

The $a_2$ homogeneous gradation which was used so far is not the 
physical gradation of the affine Toda field theory. The physical 
gradation is the so-called spin gradation, in which the rapidity 
dependence of the charges generating the quantum affine symmetry algebra 
matches the spin of the nonlocal currents from which these charges 
can be derived \cite{feldlecl,berlecl}. This gradation can be 
obtained by performing the following map on the generators:
\begin{equation}
a\rightarrow \alpha (x)a\alpha(x)^{-1},\ \alpha(x)=x^{(h_1+h_2)/6}\ .
\end{equation}
In this gradation, the evaluation representation becomes:
\begin{eqnarray}
&&\pi_{spin}(h_i)=h_i\ ,\ i=0,1,2\ ,\nn
&&\pi_{spin}(e_i)=x^{1/6}e_i\ ,\ i=1,2\ ,\nn 
&&\pi_{spin}(f_i)=x^{-1/6}f_i\ ,\ i=1,2\ ,\nn
&&\pi_{spin}(e_0)=x^{-1/2}e_0\ ,\nn 
&&\pi_{spin}(f_0)=x^{1/2}f_0\ .
\label{spineval}\end{eqnarray}
{}From (\ref{spineval}) one can read off the spin of the conserved
charges, 
using the rapidity dependence of $x$ (\ref{xtheta}). 
Note that the spins are proportional to the length of the corresponding
root. 
The $R$-matrix in the spin gradation can be obtained via
\begin{equation}
P_{12}\R_{spin}=\alpha(x_1)\otimes
\alpha(x_2)P_{12}\R_{a_2}
\alpha(x_1)^{-1}\otimes 
\alpha(x_2)^{-1}\ ,\ x=x_1/x_2\ .
\end{equation}
The charge conjugation matrix in this gradation is 
\begin{equation}
C_{spin}=\left(\matrix{
0 & 0 & 0 & 0 & 0 & 0 & 0 & 1 \cr
0 & 0 & 0 & 0 & 0 & 0 & -1 & 0 \cr
0 & 0 & 0 & 0 & 0 & -1  & 0 & 0 \cr
0 & 0 & 0 & 1 & 0 & 0 & 0 & 0 \cr
0 & 0 & 0 & 0 & 1 & 0 & 0 & 0 \cr
0 & 0 & -1 & 0 & 0 & 0 & 0 & 0\cr
0 & -1  & 0 & 0 & 0 & 0 & 0 & 0\cr
1 & 0 & 0 & 0 & 0 & 0 & 0 & 0\cr
}\right)\ .
\end{equation}

However, the $a_2$ gradation does play an important role which we 
now describe. It is well-known that imaginary coupled affine Toda field
theory 
is nonunitary and that therefore a restriction is needed to get a 
sensible physical theory. This is called the RSOS restriction and 
leads to integrable perturbations of minimal models of $W$ algebras 
\cite{Wrestr,vaysburd}. 
 
The action (\ref{atftlagr}) can be rewritten as the action of a
conformal $A_2$ Toda theory with a perturbation term:

\begin{eqnarray}
&&S=S_{A_2}+S_{pert}\ ,\nn
&&S_{A_2}=\int d^2x
\frac{1}{2}\partial_\mu\vec{\Phi}\partial_\mu\vec{\Phi}
+\frac{\lambda}{2\pi}\int d^2x \sum\limits_{j=1}^{2}
\exp\left(i\beta\frac{2}{(\vec{\alpha}_j,\vec{\alpha}_j)}
\vec{\alpha}_j\cdot\vec{\Phi}\right)\ ,\nn
&&S_{pert}=\frac{\lambda}{2\pi}\int d^2x 
\exp\left(i\beta\frac{2}{(\vec{\alpha}_0,\vec{\alpha}_0)}
\vec{\alpha}_0\cdot\vec{\Phi}\right)\ .
\end{eqnarray}

$S_{A_2}$ describes a $WA_2$-invariant conformal field theory with 
the central charge \cite{bilgerv}
\begin{equation}
c=2\left(1-12\left(\frac{\beta}{\sqrt{4\pi}}
-\frac{\sqrt{4\pi}}{\beta}\right)^2\right)\ .
\end{equation}
When 
\begin{equation}
\frac{\beta}{\sqrt{4\pi}}=\sqrt{\frac{p}{p'}}\ ,\ p,\ p'\ 
{\rm coprime}\ {\rm integers}
\end{equation}
this is just the central charge of the $(p,p')$ minimal
model of the $WA_2$ algebra, which we denote with $WA_2(p,p')$. The
field
content of the minimal model can be described by giving the spectrum of
the
primary fields. These are labelled by four integers 
$n_1,\ n_2,\ m_1,\ m_2$ and denoted $\Phi(n_1n_2|m_1m_2)$. 
To each of these fields one can associate a vector 
\begin{equation}
\vec{\beta}(n_1n_2|m_1m_2)=\sum\limits_{i=1}^2 \left( \alpha_- (1-n_i)
+\alpha_+ (1-m_i\right)\omega_i\ , 
\end{equation}
where $\omega_i$ are the fundamental weights of $A_2$ and we define
\begin{equation}
\alpha_+^2=\sqrt{\frac{p}{p'}}\ ,\ \alpha_-=-\frac{1}{\alpha_+}\ ,\ 
\alpha=\alpha_+ +\alpha_-\ .
\end{equation}
The conformal weight of the field $\Phi(n_1n_2|m_1m_2)$ is \cite{fatluk}
\begin{equation}
h(n_1n_2|m_1m_2)=\frac{1}{2}\vec{\beta}^2-\alpha\vec{\rho}\vec{\beta}\ ,
\end{equation}
where
\begin{equation}
\vec{\rho}=\sum\limits_{i=1}^2 \omega_i\ .
\end{equation}
In our case the perturbing term turns out to be the field $\Phi(11|12)$, 
which leads to an integrable perturbation \cite{mathieu}.
The weight of this field is 
\begin{equation}
h(11|12)=\frac{4}{3}\frac{p}{p'}-1\ .
\end{equation}
To get a massive field theory, we require the perturbing field to be 
a relevant one, which means that its weight is less than one. Then 
we obtain the following condition:
\begin{equation}
\frac{p}{p'}<\frac{3}{2}\ {\rm or}\ \beta^2<6\pi
\label{relevcond}\end{equation}
This condition is always satisfied for unitary minimal models of $WA_2$ 
($|p-p'|=1$). 

For the above choice of $\beta$ the parameter $q$ becomes a root of
unity. We therefore expect that there exists a consistent restriction 
of the affine Toda field theory to the perturbed minimal model, 
which is described 
by the corresponding restriction of the representation theory of 
${\cal U}_q(a_2)$ at this value of $q$. This is the point where 
the $a_2$ homogeneous gradation enters the game, since in this 
gradation the generators of ${\cal U}_q(a_2)$ are Lorentz invariant 
and therefore the space of states can be restricted under the 
action of ${\cal U}_q(a_2)$. In fact, changing the gradation can 
be seen as redefining the energy-momentum tensor by a Feigin-Fuchs 
term, which is a crucial step before RSOS restriction \cite{smirresh}. 
One must have a new energy-momentum tensor commuting with the 
${\cal U}_q(a_2)$ charges in order to get a consistent relativistic 
theory after the restriction.

Eqn. (\ref{relevcond}) can also be viewed as the condition 
$\xi >0$. In fact, using an argument similar to \cite{g2},   
for $\xi <0$ we expect that the theory becomes trivial in the infrared 
in analogy with sine-Gordon theory. Then the unrestricted theory  
describes two free massless fields and the RSOS restriction must 
coincide with the free field realization of $WA_2$ minimal models 
\cite{fatluk}.

\section{Conclusion}

In this paper we have constructed the $R$-matrix of the $\qda$ algebra 
in the $8$-dimensional fundamental representation. The construction 
is made difficult by the occurrence of multiplicities in the tensor 
product. It can be seen that the result is essentially nondiagonal 
in the multiplicity space (i.e. in the space of multiplicity labels 
of the subspace ${\bf 8}+{\bf 8}$) which means that it can not be 
diagonalized in a basis independent of the spectral parameter $x$. 
This is analogous to a phenomenon observed by the author in connection 
with the fundamental soliton $S$-matrix in the $\Phi_{(1,5)}$
perturbation 
of the Virasoro minimal model $Vir(3,16)$ \cite{rsos}. Therefore 
there seems to be no way to fit this $R$-matrix into the usual scheme 
of the tensor product graph method, although the TPG applies in the 
subspace which is multiplicity-free. It was also pointed out that 
the degeneration of the $R$-matrix at its pole singularities follows 
closely the pattern observed in the case of  
${\cal U}_q\left(g_2^{(1)}\right)$ in \cite{g2}. 

Using the $R$-matrix, the scattering amplitudes for the fundamental 
soliton of imaginary coupled $g_2^{(1)}$ affine Toda field theory 
were constructed in the $a_2$ gradation. The scattering amplitude 
of the first breather was computed and shown to correspond to 
the $S$-matrix of the second particle of the real coupling theory. 
It is proposed that the first particle must come from the other 
type of solitonic multiplets, but unfortunately it is not possible 
to verify this statement due to technical complications. This is 
an open question for future investigation.

Another interesting open question is the application of an RSOS 
restriction procedure to this $S$-matrix to get the scattering 
amplitudes of $\Phi(11|12)$ perturbations of $WA_2$ minimal models. 
These models are more interesting than the $\Phi(11|14)$ perturbations 
studied in \cite{g2} because they include unitary integrable quantum
field 
theories as well. 

The solution of these questions would help us in understanding
more of the physics behind affine Toda field theories. Some of these 
issues were already mentioned in \cite{g2}. In particular, it would be 
interesting to see more evidence for the validity of the approach 
based on the quantum symmetry algebra. It is not obvious that this 
approach is correct and the main issue is that classically the 
soliton solutions fail to fill up the affine algebra multiplets
completely 
\cite{mcghee}, while the quantum symmetry approach takes this for
granted 
at the quantum level. It would also be interesting to see how a unitary 
restriction can emerge from a theory which is strongly nonunitary, 
taking into account especially the instability of classical solitonic 
solutions \cite{khastgir} and the strong unitarity violation at the 
quantum level that may persist even after an RSOS restriction
\cite{rsos}.
\vspace{.5in}
\begin{center}
{\bf Acknowledgements}
\end{center}

I wish to thank SISSA and especially G. Mussardo for their kind 
hospitality. I would also like to acknowledge useful discussions with 
G.M.T. Watts and N.J. Mackay, and the valuable comments from A.
Babichenko 
in connection with my previous paper.

\end{document}